\documentclass[prd,aps,showpacs,floatfix,twocolumn]{revtex4}

\usepackage{amsmath,amsfonts}
\usepackage[dvips]{graphicx}

\newcommand{\ud}{\mathrm{d}}

\begin{document}

\title{Coupling constant constraints in a nonminimally coupled phantom cosmology}

\author{Marek Szyd{\l}owski}
\email{uoszydlo@cyf-kr.edu.pl}
\affiliation{Astronomical Observatory, Jagiellonian University,
Orla 171, 30-244 Krak{\'o}w, Poland}
\affiliation{Mark Kac Complex Systems Research Centre, Jagiellonian University,
Reymonta 4, 30-059 Krak{\'o}w, Poland}
\author{Orest Hrycyna} 
\email{hrycyna@kul.lublin.pl}
\affiliation{Department of Theoretical Physics, Faculty of Philosophy, 
The John Paul II Catholic University of Lublin, Al. Rac{\l}awickie 14, 20-950
Lublin, Poland}
\author{Aleksandra Kurek}
\email{alex@oa.uj.edu.pl}
\affiliation{Astronomical Observatory, Jagiellonian University,
Orla 171, 30-244 Krak{\'o}w, Poland}

\date{\today}

\begin{abstract}
In the paper we investigate observational constraints on coupling to gravity 
constant parameter $\xi$ using distant supernovae SNIa data, baryon 
oscillation peak (BOP), the cosmic microwave background radiation
(CMBR) shift parameter, and $H(z)$ data set. We 
estimate the value of this parameter to constrain the extended quintessence 
models with nonminimally coupled to gravity phantom scalar field. 
The combined analysis of observational data favors a value of $\xi$ which lies 
in close neighborhood of the conformal coupling. While our estimations are model dependent
they give rise to a indirect bound on the Equivalence Principle.
\end{abstract}


\maketitle

At the present, scalar fields play the crucial role in modern cosmology. In the 
inflationary scenario they generate an exponential rate of evolution of the
universe as well as a density fluctuations due to vacuum energy. 

Observations of distant supernovae support the cosmological constant term.
But two problems emerge in this context. Namely, the  fine
tuning and cosmic coincidence problems.
The lack of some fundamental mechanism which sets the cosmological
constant almost zero is called the cosmological constant problem. The second
problem called ``cosmic conundrum'' is the question why the energy densities
of both dark energy and dark matter are nearly equal at the present epoch.
One of the solutions to this problem offers the idea of quintessence 
\cite{Wetterich:1987fm, Ratra:1987rm} which is a
version of the time varying cosmological constant conception.

All these models base on the assumption that there is a minimal coupling of 
scalar field to gravity ($\xi=0$). This \textit{a priori} assumption requires 
some justification. There are many theoretical arguments suggesting that the 
non-minimal coupling (NC for short) should be considered. The nonzero $\xi$ 
comes from quantum corrections \cite{Birrell:1979ip}, renormalization of 
classical theory that shifts it to one with nonzero $\xi$ \cite{Callan:1970ze}, 
in relativity the value of $\xi=1/6$ (conformal coupling) is distinguished. 
Only in this case the Einstein equivalence principle is not violated (for 
details see \cite{Sonego:1993fw}). 

It is expected also that the value of coupling constant $\xi$ should be fixed 
by the physics of the problem and there is no free parameters, and then we obtain
answer to the question ``what is the value of $\xi$?'' \cite{Faraoni:2000gx}.
However, the answer to this question differs according to the theory of the
scalar field employed. For example, in Einstein gravity with the polynomial
quartic potential function and back-reaction, the value of NC was found
\cite{Parker:1985kc} or a Higgs field in the standard model gives value of NC
non-positive or larger (or equal) than $1/6$ \cite{Hosotani:1985at}.

The main goal of 
this paper is a statistical estimation of the coupling parameter from the
astronomical observations. For this aim we consider the spatially flat FRW model
where the source of gravity is a noninteracting mixture of dust matter and
non-minimally coupled phantom scalar field.

We treat $\xi$ as a free parameter in the model 
and we are looking for constraints on it's value from
observational cosmology. For simplicity it is assumed a simple and natural form
of the potential function of the scalar field. For the vanishing coupling 
constant (minimal coupling) this potential corresponds to chaotic inflation
\cite{Linde:1983gd}. This paradigm of inflation can be extended by including NC
\cite{Faraoni:2000gx}. Also the scalar field non-minimally coupled to gravity
are simple, non-exotic models of phantoms give rise to superacceleration
\cite{Faraoni:2001tq}.

Tsujikawa and Gumjudpai \cite{Tsujikawa:2004my} also investigated constraints 
on the NC parameter from the CMB observations. Recently Jankiewicz and 
Kephart \cite{Jankiewicz:2005tm} applied the method of WKB approximation to 
study the NC in the FRW universe with the scalar field and found the best fit
value the NC parameter very close to $1/6$. Futamase and Maeda 
\cite{Futamase:1987ua} in the case of the 
quadratic potential of the inflaton field found that modulus of NC  
less than $10^{-3}$ leads to sufficient inflation (see also 
\cite{Tsujikawa:2000tm} for other constraints). 

In order to determine the NC parameter we use a recent expansion history of the
universe rather than early stages of its evolution (inflation). In estimation of
the parameter $\xi$ we do not assume any arbitrary parametrization for the
equation of state parameter $w(z)$, but we derive its exact form directly from
the model (dynamics). While this equation has in general very complex form, 
in practice its simple approximations are used. If we expand the 
$w(z)$ relation in the Taylor series with the respect to the redshift $z$ then we 
recover the well known parametrization in the linear form of the redshift or 
the scale factor. All the coefficients in the series depend on the parameter 
$\xi$ and initial conditions. Then after constraining initial conditions from 
the observational data we can obtain corresponding limits on $\xi$.

We assume the flat model with the FRW geometry with the signature $(-,+,+,+)$
and a source of gravity is the phantom scalar field
$\psi$ with an arbitrary coupling constant $\xi$. The dynamics is governed
by the action
\begin{equation}
S=\frac{1}{2}\int \ud^{4}x \sqrt{-g}\Big(m_{p}^{2}R +
(g^{\mu\nu}\psi,_{\mu}\psi,_{\nu} + \xi R\psi^{2} - m^{2}\psi^{2})\Big)
\label{eq:3} 
\end{equation}
where $m_{p}^{2}=(8\pi G)^{-1}$; for simplicity and without lost of generality
we assume $4\pi G/3=1$ and $m^{2}=1$.

In our paper \cite{Hrycyna:2007gd} we studied generic features of the evolutionary
paths of the flat FRW model with phantom scalar field non-minimally coupled to
gravity. We reduced dynamics of the model to the simple case of autonomous
dynamical system on the invariant submanifold $(\psi,\psi')$ (a prime denotes
differentiation with respect to the natural logarithm of the scale factor) and
we founded that principally there is one asymptotic state, which corresponds to
the critical point in the phase space $\psi_{0}=\pm1/\sqrt{6\xi}$ and
$\psi_{0}'=0$. This critical point is also the de Sitter state ($w=-1$). There 
are two types of scenario leading to this Lambda state (depending on the 
value of $\xi$), through
\begin{itemize}
\item[1.]{the monotonic evolution toward the critical point of a node
type for $0<\xi\le3/25$;}
\item[2.]{the damping oscillations around the critical point of a focus type
for $3/25<\xi<1/3$.}
\end{itemize}

In both evolutionary scenarios we obtained linearized solutions of the dynamical
system in the vicinity of the critical point. Using relation of the
equation of state parameter on the dynamical variables $\psi$ and $\psi'$ 
we were able to derive exact forms of $w(z)$ relations in both scenarios 
(see \cite{Hrycyna:2007gd} for details).

Now, we present method of estimation model parameters from 
combined analysis of the parameters $w_0$, $w_1$ and $w_2$ from 
the combined astronomical data for three parametrization of $w(z)$ (see Table \ref{tab:1}). 

\begin{table*}
\caption{\label{tab:1} Estimations of expansion coefficients for three different 
parametrization of $w(z)$.}
\begin{ruledtabular}
\begin{tabular}{c|c|ccc}
Case & parametrization & $w_{0}$ & $w_{1}$ & $w_{2}$ \\
\hline
$(1)$ & $w(z)=w_{0}+w_{1}z+w_{2}z^{2}$ & $-1.29\pm0.18$ & $1.50\pm1.03$ & $-0.57\pm0.75$
\\
$(2)$ & $w(z)=w_{0}+w_{1}\frac{z}{1+z}+w_{2}(\frac{z}{1+z})^{2}$ & $-1.38\pm0.13$ &
$2.75\pm0.87$ & $-2.13\pm1.33$ \\
$(3)$ & $w(z)=w_{0}+w_{1}\ln{(1+z)}+w_{2}(\ln{(1+z)})^{2}$ & $-1.34\pm0.24$ &
$2.16\pm1.71$ & $-1.25\pm1.73$
\end{tabular}
\end{ruledtabular}
\end{table*}

\begin{table}
\caption{\label{tab:2} Solutions to the system of equations 
$w^{0}_{X}$, $w^{1}_{X}$ and $w^{2}_{X}$
with values of $w_{0}$, $w_{1}$ and $w_{2}$ from Table \ref{tab:1}.}
\begin{ruledtabular}
\begin{tabular}{c|ccc|c}
Case & $x_{0}$ & $y_{0}$ & $y_{0}/x_{0}$ & $\xi$ \\
\hline
$(1)$ & $7.45\pm7.05$ & $-8.55\pm7.77$ & $-1.15\pm2.13$ & $0.14\pm0.03$\\
$(2)$ & $6.44\pm2.75$ & $-8.22\pm3.11$ & $-1.28\pm1.03$ & $0.15\pm0.02$\\
$(3)$ & $6.53\pm8.77$ & $-8.09\pm10.12$ & $-1.24\pm3.21$ & $0.15\pm0.05$
\end{tabular}
\end{ruledtabular}
\end{table}

To constrain the unknown values of model parameters we used the the set of 
$N_1=192$ SNIa data, CMBR shift parameter, measurement of the parameter $A$ 
coming from the SDSS luminous red galaxies and $N_2=9$ observational $H(z)$ 
data. The likelihood function for such data sets has the following form
\begin{equation}
\mathcal{L}(\bar{\theta}) \propto \exp\left[-\frac{1}{2}\chi^2(\bar{\theta})\right],
\end{equation}
where 
$\bar{\theta}=(H_0,\Omega_{m,0},w_0,w_1,w_2)$ and $\chi^2(\bar{\theta})=
\chi_{SN}^2(\bar{\theta})+\chi_{R}^2(\bar{\theta})+\chi_{A}^2(\bar{\theta})+
\chi_{H}^2(\bar{\theta})$.

Let us consider quantities from the definition of function $\chi^2$.
Using the SNIa data \cite{Riess:2006fw, Davis:2007na, WoodVasey:2007jb} we base
on 
the standard relation between the apparent magnitude $m$ and luminosity 
distance $d_L$: $m-M=5\log _{10} D_L + \mathcal{M}$, where $M$ is the absolute 
magnitude of SNIa, $\mathcal{M}=-5\log _{10} H_0 +25$ and $D_L=H_0 d_L$. The 
luminosity distance depends on the cosmological model considered and with the 
assumption that $k=0$ is given by $d_{L}=(1+z)c\int_{0}^{z} \frac{d z'}{H(z')}$.
The $\chi^2_{\mathrm{SN}}$ function has the following form
\begin{equation}
 \chi_{\mathrm{SN}}^{2}(\bar{\theta})=\sum _{i=1}^{N_1} 
\left (\frac{\mu_i^{\mathrm{obs}}-\mu_i^{th}}{\sigma_i} \right )^2,
\end{equation}
where $\mu_{i}^{\mathrm{obs}}=m_{i}-M$, $\mu_{i}^{th}=5\log_{10}D_{Li} + \mathcal{M}$.

As we have written before we also used constraints which comes from the so 
called CMBR shift parameter which is defined as 
$R^{th}=\sqrt{\Omega_{\mathrm{m},0}}\int_{0}^{z_{\mathrm{dec}}}\frac{H_0}{H(z)}dz$. 
The $\chi_{R}^2$ has the following form
\begin{equation}
\chi ^2_{R}(\bar{\theta}) =\left( \frac{R^{obs}-R^{th}}{\sigma_R} \right)^2,
\end{equation}
where $R^{{obs}}=1.70 \pm 0.03$ for $z_{\mathrm{dec}}=1089$ \cite{Spergel:2006hy, Wang:2006ts}.

We also used constrains coming from the SDSS luminous red galaxies measurement 
of $A$ parameter ($A^{\mathrm{obs}}=0.469 \pm 0.017$ for $z_{A}=0.35$) 
\cite{Eisenstein:2005su}, which is related to the baryon acoustic oscillation 
peak and defined in the following way $A^{\mathrm{th}}=\sqrt{\Omega_{\mathrm{m},0}} 
\left (\frac{H(z_A)}{H_{0}} \right ) ^{-\frac{1}{3}} \left [ \frac{1}{z_{A}} 
\int_{0}^{z_{A}}\frac {H_0}{H(z)}d z\right]^{\frac{2}{3}}.$
The $\chi^2_A$ function has the following form 
\begin{equation}
\chi ^2_{A}(\bar{\theta})=\left( \frac{A^{th}-A^{\mathrm{obs}}}{\sigma_{A}}\right)^2.
\end{equation}

Finally we add constraints coming from observational $H(z)$ data ($N=9$) 
\cite{Simon:2004tf, Samushia:2006fx, Wei:2006ut}. This data based on the 
differential ages $(\ud t/\ud z)$ of the passively evolving galaxies which 
allow to estimate the relation 
$H(z) \equiv \dot{a}/a =-(1+z)^{-1} \ud z/\ud t$. 
The function $\chi^2_H$ has the following form  
\begin{equation}
\chi ^2_{H}(\bar{\theta})= \sum_{i=1}^{N_2} \left( \frac{H(z_i) -H_i(z_i)}{\sigma_i ^2}\right)^2.
\end{equation}

After the marginalization likelihood function over the parameter $H_0$ in the 
range $<60,80>$ the values of model parameters $(\Omega_{\mathrm{m},0},w_0,w_1,w_2)$ 
were obtained via the $\chi^2$ minimization procedure and are gathered in 
Table \ref{tab:1}. There are also the $1\sigma$ uncertainties which correspond 
to the parameter value where the likelihood function differ from the best fit 
likelihood by a factor of $e^{\frac{1}{2}}$.

For every fitting parametrization we obtain a set of three numbers 
$(w_{0},w_{1},w_{2})$ (see Table \ref{tab:1}). Then basing on expansion 
in the Taylor series of the exact forms of EoS functions up to second order term
$(w^{0}_{X},w^{1}_{X},w^{2}_{X})$,
we can solve the 
system of three equation for three values $\xi$, $x_{0}$ i $y_{0}$.
The errors of $\Delta x_{0}$, $\Delta y_{0}$ and $\Delta \xi$ are calculated in
a standard way. 
From Table \ref{tab:2} we can see that the
parameter 
$\xi$ is closed to the its conformal value $1/6$. Therefore the combined 
analysis of observational data favors a value of $\xi$ which lies in close 
neighborhood of the conformal coupling. The second conclusion is that the 
oscillatory scenario is favored. In the case of parametrization (2) we can 
exclude the monotonic scenario on the $1\sigma$ confidence level. 

Finally we consider parameters $w_0$, $w_1$, $w_2$ as functions of parameters 
$\xi$, $x_0$, $y_0$. 
The values of model parameters $\Omega_{\mathrm{m},0}$, $\xi$, $x_0$, $y_0$ 
obtained via the $\chi^2$ minimization procedure together with $1\sigma$ 
errors are gathered in Table \ref{tab:3}. 

The initial condition $(x_0, y_0)$ defines the initial rate of change of
scalar field $(\ln(\psi_{i}-\psi_{0}))' = \psi_{i}'/(\psi_{i}-\psi_{0}) =
y_{0}/x_{0}=(\ln x_{0})'$. This value is negative and is
presented in the third column of Table \ref{tab:2} and in the fourth column of
Table \ref{tab:3}.

\begin{table*}
\caption{The values of directly estimated parameters $\Omega_{\mathrm{m},0}$, $\xi$, 
$x_0$, $y_0$ after substituting $w^{0}_{X}$, $w^{1}_{X}$, $w^{2}_{X}$ into the $H(z)$ formula 
for three different parametrization of $w(z)$.}
\begin{ruledtabular}
\begin{tabular}{c|cccc|c}
Case & $\Omega_{\mathrm{m},0}$ & $x_{0}$ & $y_{0}$ &$y_{0}/x_{0}$& $\xi$ \\
\hline
$(1)$ & $0.27\pm0.02$ & $7.33\pm0.90$ & $-8.43\pm0.75$ &$-1.15\pm0.24$& $0.14\pm0.01$ \\
$(2)$ & $0.27\pm0.02$ & $6.41\pm0.32$ & $-8.18\pm0.32$ &$-1.28\pm0.11$& $0.15\pm0.01$ \\
$(3)$ & $0.27\pm0.02$ & $6.60\pm0.72$ & $-8.16\pm0.81$ &$-1.24\pm0.26$& $0.15\pm0.01$
\end{tabular}
\end{ruledtabular}
\label{tab:3}
\end{table*}

We can see that in this case the values of parameters do not differ from the values 
obtained by indirect method via the estimation of $w_0$, $w_1$, $w_2$ but 
here the uncertainties of the parameter values are smaller. In Table \ref{tab:4} we
gathered the values of the parameter of non-minimal coupling $\xi$ estimated via
direct method together with $1\sigma$, $2\sigma$ and $3\sigma$ confidence
levels. The value of $\xi = 1/6$ is consistent with our results at a $3\sigma$ 
confidence level for
parametrization (1) and at a $2\sigma$ confidence level for parametrization (2)
and (3). As one can conclude the oscillatory scenario is favored, moreover, we
can exclude the monotonic scenario at a $2\sigma$ confidence level for the
parametrization (1) and (3) and at a $3\sigma$ confidence level for
parametrization (2).

\begin{table}
\caption{The values of the parameter of non-minimal coupling $\xi$ estimated via
direct method together with $1\sigma$, $2\sigma$ and $3\sigma$ confidence
levels. In the
cases (1) and (3) we can exclude the monotonic scenario at a $2\sigma$
and in the case (2) at a $3\sigma$ level.}
\begin{ruledtabular}
\begin{tabular}{c|cccc}
Case & $\xi$ & $1\sigma$ & $2\sigma$ & $3\sigma$ \\
\hline
$(1)$ & $0.14$ & $0.01$ & $0.02$ & $0.04$ \\
$(2)$ & $0.15$ & $0.01$ & $0.02$ & $0.03$ \\
$(3)$ & $0.15$ & $0.01$ & $0.02$ & $0.05$ 
\end{tabular}
\end{ruledtabular}
\label{tab:4}
\end{table}

While the standard quintessence idea is most
popular one, scalar field cosmology with minimally coupled scalar field require
existence of a special conditions for a tracker solution and potential functions
of the scalar field \cite{Steinhardt:1999nw}. The standard quintessence
evolutionary scenario has been extended by introduction of the non-minimal
coupling between gravitation and scalar field. Hence the cosmology with such a
fields contains an additional parameter which should be determined from a
observational data. We demonstrate that astronomical data and cosmography may be
very useful in estimation of this free parameter of a model.

In our approach the effects of non vanishing coupling constant are dynamically
equivalent to the effects of substantial dark energy characterized by the 
coefficient of the equation of state. The dynamics with positive value of the 
NC parameter admits two types of evolutionary scenarios: the monotonic evolution and 
damping oscillations around the stationary de Sitter state. Given the 
estimated value of the parameter $\xi$ we found that the oscillating scenario 
is favored. 

We expanded the exact form of the EoS parameter up to second order term and 
then constrained the effective form of the EoS which determines directly the 
value of the NC parameter. As a result performed combined analysis
we obtain that value of $\xi$ from neighborhood of conformal coupling is
favored by observational astronomical data, i.e. SN Ia data, Baryon
Oscillation Peak, CMBR shift and $H(z)$ data. It was demonstrated that value of
estimated NC parameter does not depend on used expansion formula up to the 
second order although errors can be different. 

The General Relativity in based on the Einstein Equivalence Principle, which
includes Weak Equivalence Principle, Local Lorentz Invariance and Local Position
Invariance. Therefore if
this universality condition is valid then gravity can be described in terms of
Riemannian geometry. Our estimation of coupling constant is based on the FRW
model filled by non-minimally coupled phantom scalar field 
but one should also note the existence of the direct constraints on WEP.
Then deviation value of coupling constant from the conformal coupling can
measure the deviation from the WEP and the GRG in the Einstein formulation.
Our estimation, based on a model with the non-minimally coupled phantom scalar
field, gives the deviation from the WEP $\Delta\xi/\xi\simeq10^{-1}$.
On the other hand the string theory suggest the existence of strength
scalar field (dilations) whose couplings to matter can violate the WEP
\cite{Damour:2001fn}. This provides a new motivation for high-precision
experiments verifying the universality of free fall to the
$10^{-12}$ level \cite{Su:1994gu, Damour:1996xt, Fuzfa:2006pn}. The lunar Laser
Ranging experiment has also verified that the Moon and the Earth fall with the
same acceleration toward the Sun to better than one part
in $10^{12}$ precision \cite{Dickey:1994}. Also observed limits on evolution of
the binary pulsar BO655+64 orbit provide new bounds on the violation of SEP
\cite{Arzoumanian:2002xu} (for the description of current and future projects
for the improvement of the accuracy of the experiments as compared experiments
on ground see \cite{Lammerzahl:2004tc}).

This work has been supported by the Marie Curie Actions
Transfer of Knowledge project COCOS (contract MTKD-CT-2004-517186).


\begin{thebibliography}{32}
\expandafter\ifx\csname natexlab\endcsname\relax\def\natexlab#1{#1}\fi
\expandafter\ifx\csname bibnamefont\endcsname\relax
  \def\bibnamefont#1{#1}\fi
\expandafter\ifx\csname bibfnamefont\endcsname\relax
  \def\bibfnamefont#1{#1}\fi
\expandafter\ifx\csname citenamefont\endcsname\relax
  \def\citenamefont#1{#1}\fi
\expandafter\ifx\csname url\endcsname\relax
  \def\url#1{\texttt{#1}}\fi
\expandafter\ifx\csname urlprefix\endcsname\relax\def\urlprefix{URL }\fi
\providecommand{\bibinfo}[2]{#2}
\providecommand{\eprint}[2][]{\url{#2}}

\bibitem[{\citenamefont{Wetterich}(1988)}]{Wetterich:1987fm}
\bibinfo{author}{\bibfnamefont{C.}~\bibnamefont{Wetterich}},
  \bibinfo{journal}{Nucl. Phys.} \textbf{\bibinfo{volume}{B302}},
  \bibinfo{pages}{668} (\bibinfo{year}{1988}).

\bibitem[{\citenamefont{Ratra and Peebles}(1988)}]{Ratra:1987rm}
\bibinfo{author}{\bibfnamefont{B.}~\bibnamefont{Ratra}} \bibnamefont{and}
  \bibinfo{author}{\bibfnamefont{P.~J.~E.} \bibnamefont{Peebles}},
  \bibinfo{journal}{Phys. Rev.} \textbf{\bibinfo{volume}{D37}},
  \bibinfo{pages}{3406} (\bibinfo{year}{1988}).

\bibitem[{\citenamefont{Birrell and Davies}(1980)}]{Birrell:1979ip}
\bibinfo{author}{\bibfnamefont{N.~D.} \bibnamefont{Birrell}} \bibnamefont{and}
  \bibinfo{author}{\bibfnamefont{P.~C.~W.} \bibnamefont{Davies}},
  \bibinfo{journal}{Phys. Rev.} \textbf{\bibinfo{volume}{D22}},
  \bibinfo{pages}{322} (\bibinfo{year}{1980}).

\bibitem[{\citenamefont{Callan~Jr. et~al.}(1970)\citenamefont{Callan~Jr.,
  Coleman, and Jackiw}}]{Callan:1970ze}
\bibinfo{author}{\bibfnamefont{C.~G.} \bibnamefont{Callan~Jr.}},
  \bibinfo{author}{\bibfnamefont{S.~R.} \bibnamefont{Coleman}},
  \bibnamefont{and} \bibinfo{author}{\bibfnamefont{R.}~\bibnamefont{Jackiw}},
  \bibinfo{journal}{Ann. Phys.} \textbf{\bibinfo{volume}{59}},
  \bibinfo{pages}{42} (\bibinfo{year}{1970}).

\bibitem[{\citenamefont{Sonego and Faraoni}(1993)}]{Sonego:1993fw}
\bibinfo{author}{\bibfnamefont{S.}~\bibnamefont{Sonego}} \bibnamefont{and}
  \bibinfo{author}{\bibfnamefont{V.}~\bibnamefont{Faraoni}},
  \bibinfo{journal}{Class. Quant. Grav.} \textbf{\bibinfo{volume}{10}},
  \bibinfo{pages}{1185} (\bibinfo{year}{1993}).

\bibitem[{\citenamefont{Faraoni}(2001)}]{Faraoni:2000gx}
\bibinfo{author}{\bibfnamefont{V.}~\bibnamefont{Faraoni}},
  \bibinfo{journal}{Int. J. Theor. Phys.} \textbf{\bibinfo{volume}{40}},
  \bibinfo{pages}{2259} (\bibinfo{year}{2001}), \eprint{arXiv:hep-th/0009053}.

\bibitem[{\citenamefont{Parker and Toms}(1985)}]{Parker:1985kc}
\bibinfo{author}{\bibfnamefont{L.}~\bibnamefont{Parker}} \bibnamefont{and}
  \bibinfo{author}{\bibfnamefont{D.~J.} \bibnamefont{Toms}},
  \bibinfo{journal}{Phys. Rev.} \textbf{\bibinfo{volume}{D32}},
  \bibinfo{pages}{1409} (\bibinfo{year}{1985}).

\bibitem[{\citenamefont{Hosotani}(1985)}]{Hosotani:1985at}
\bibinfo{author}{\bibfnamefont{Y.}~\bibnamefont{Hosotani}},
  \bibinfo{journal}{Phys. Rev.} \textbf{\bibinfo{volume}{D32}},
  \bibinfo{pages}{1949} (\bibinfo{year}{1985}).

\bibitem[{\citenamefont{Linde}(1983)}]{Linde:1983gd}
\bibinfo{author}{\bibfnamefont{A.~D.} \bibnamefont{Linde}},
  \bibinfo{journal}{Phys. Lett.} \textbf{\bibinfo{volume}{B129}},
  \bibinfo{pages}{177} (\bibinfo{year}{1983}).

\bibitem[{\citenamefont{Faraoni}(2002)}]{Faraoni:2001tq}
\bibinfo{author}{\bibfnamefont{V.}~\bibnamefont{Faraoni}},
  \bibinfo{journal}{Int. J. Mod. Phys.} \textbf{\bibinfo{volume}{D11}},
  \bibinfo{pages}{471} (\bibinfo{year}{2002}), \eprint{arXiv:astro-ph/0110067}.

\bibitem[{\citenamefont{Tsujikawa and Gumjudpai}(2004)}]{Tsujikawa:2004my}
\bibinfo{author}{\bibfnamefont{S.}~\bibnamefont{Tsujikawa}} \bibnamefont{and}
  \bibinfo{author}{\bibfnamefont{B.}~\bibnamefont{Gumjudpai}},
  \bibinfo{journal}{Phys. Rev.} \textbf{\bibinfo{volume}{D69}},
  \bibinfo{pages}{123523} (\bibinfo{year}{2004}),
  \eprint{arXiv:astro-ph/0402185}.

\bibitem[{\citenamefont{Jankiewicz and Kephart}(2006)}]{Jankiewicz:2005tm}
\bibinfo{author}{\bibfnamefont{M.}~\bibnamefont{Jankiewicz}} \bibnamefont{and}
  \bibinfo{author}{\bibfnamefont{T.~W.} \bibnamefont{Kephart}},
  \bibinfo{journal}{Phys. Rev.} \textbf{\bibinfo{volume}{D73}},
  \bibinfo{pages}{123514} (\bibinfo{year}{2006}),
  \eprint{arXiv:hep-ph/0510009}.

\bibitem[{\citenamefont{Futamase and Maeda}(1989)}]{Futamase:1987ua}
\bibinfo{author}{\bibfnamefont{T.}~\bibnamefont{Futamase}} \bibnamefont{and}
  \bibinfo{author}{\bibfnamefont{K.-i.} \bibnamefont{Maeda}},
  \bibinfo{journal}{Phys. Rev.} \textbf{\bibinfo{volume}{D39}},
  \bibinfo{pages}{399} (\bibinfo{year}{1989}).

\bibitem[{\citenamefont{Tsujikawa}(2000)}]{Tsujikawa:2000tm}
\bibinfo{author}{\bibfnamefont{S.}~\bibnamefont{Tsujikawa}},
  \bibinfo{journal}{Phys. Rev.} \textbf{\bibinfo{volume}{D62}},
  \bibinfo{pages}{043512} (\bibinfo{year}{2000}),
  \eprint{arXiv:hep-ph/0004088}.

\bibitem[{\citenamefont{Hrycyna and Szydlowski}(2007)}]{Hrycyna:2007gd}
\bibinfo{author}{\bibfnamefont{O.}~\bibnamefont{Hrycyna}} \bibnamefont{and}
  \bibinfo{author}{\bibfnamefont{M.}~\bibnamefont{Szydlowski}},
  \bibinfo{journal}{Phys. Rev.} \textbf{\bibinfo{volume}{D76}},
  \bibinfo{pages}{123510} (\bibinfo{year}{2007}), \eprint{arXiv:0707.4471
  [hep-th]}.

\bibitem[{\citenamefont{Davis et~al.}(2007)\citenamefont{Davis, Mortsell,
  Sollerman, Becker, Blondin, Challis, Clocchiatti, Filippenko, Foley,
  Garnavich et~al.}}]{Davis:2007na}
\bibinfo{author}{\bibfnamefont{T.~M.} \bibnamefont{Davis}},
  \bibinfo{author}{\bibfnamefont{E.}~\bibnamefont{Mortsell}},
  \bibinfo{author}{\bibfnamefont{J.}~\bibnamefont{Sollerman}},
  \bibnamefont{et~al.}, \bibinfo{journal}{Astrophys. J.}
  \textbf{\bibinfo{volume}{666}}, \bibinfo{pages}{716} (\bibinfo{year}{2007}),
  \eprint{arXiv:astro-ph/0701510}.

\bibitem[{\citenamefont{Riess et~al.}(2007)\citenamefont{Riess, Strolger,
  Casertano, Ferguson, Mobasher, Gold, Challis, Filippenko, Jha, Li
  et~al.}}]{Riess:2006fw}
\bibinfo{author}{\bibfnamefont{A.~G.} \bibnamefont{Riess}},
  \bibinfo{author}{\bibfnamefont{L.-G.} \bibnamefont{Strolger}},
  \bibinfo{author}{\bibfnamefont{S.}~\bibnamefont{Casertano}},
  \bibnamefont{et~al.},
  \bibinfo{journal}{Astrophys. J.} \textbf{\bibinfo{volume}{659}},
  \bibinfo{pages}{98} (\bibinfo{year}{2007}), \eprint{arXiv:astro-ph/0611572}.

\bibitem[{\citenamefont{Wood-Vasey et~al.}(2007)\citenamefont{Wood-Vasey,
  Miknaitis, Stubbs, Jha, Riess, Garnavich, Kirshner, Aguilera, Becker,
  Blackman et~al.}}]{WoodVasey:2007jb}
\bibinfo{author}{\bibfnamefont{W.~M.} \bibnamefont{Wood-Vasey}},
  \bibinfo{author}{\bibfnamefont{G.}~\bibnamefont{Miknaitis}},
  \bibinfo{author}{\bibfnamefont{C.~W.} \bibnamefont{Stubbs}},
  \bibnamefont{et~al.} (\bibinfo{collaboration}{ESSENCE Collaboration}),
  \bibinfo{journal}{Astrophys. J.} \textbf{\bibinfo{volume}{666}},
  \bibinfo{pages}{694} (\bibinfo{year}{2007}), \eprint{arXiv:astro-ph/0701041}.

\bibitem[{\citenamefont{Spergel et~al.}(2007)\citenamefont{Spergel, Bean, Doré,
  Nolta, Bennett, Dunkley, Hinshaw, Jarosik, Komatsu, Page
  et~al.}}]{Spergel:2006hy}
\bibinfo{author}{\bibfnamefont{D.~N.} \bibnamefont{Spergel}},
  \bibinfo{author}{\bibfnamefont{R.}~\bibnamefont{Bean}},
  \bibinfo{author}{\bibfnamefont{O.}~\bibnamefont{Doré}}, \bibnamefont{et~al.}
  (\bibinfo{collaboration}{WMAP Collaboration}), \bibinfo{journal}{Astrophys.
  J. Suppl.} \textbf{\bibinfo{volume}{170}}, \bibinfo{pages}{377}
  (\bibinfo{year}{2007}), \eprint{arXiv:astro-ph/0603449}.

\bibitem[{\citenamefont{Wang and Mukherjee}(2006)}]{Wang:2006ts}
\bibinfo{author}{\bibfnamefont{Y.}~\bibnamefont{Wang}} \bibnamefont{and}
  \bibinfo{author}{\bibfnamefont{P.}~\bibnamefont{Mukherjee}},
  \bibinfo{journal}{Astrophys. J.} \textbf{\bibinfo{volume}{650}},
  \bibinfo{pages}{1} (\bibinfo{year}{2006}), \eprint{arXiv:astro-ph/0604051}.

\bibitem[{\citenamefont{Eisenstein et~al.}(2005)\citenamefont{Eisenstein,
  Zehavi, Hogg, Scoccimarro, Blanton, Nichol, Scranton, Seo, Tegmark, Zheng
  et~al.}}]{Eisenstein:2005su}
\bibinfo{author}{\bibfnamefont{D.~J.} \bibnamefont{Eisenstein}},
  \bibinfo{author}{\bibfnamefont{I.}~\bibnamefont{Zehavi}},
  \bibinfo{author}{\bibfnamefont{D.~W.} \bibnamefont{Hogg}},
  \bibnamefont{et~al.}
  (\bibinfo{collaboration}{SDSS Collaboration}), \bibinfo{journal}{Astrophys.
  J.} \textbf{\bibinfo{volume}{633}}, \bibinfo{pages}{560}
  (\bibinfo{year}{2005}), \eprint{arXiv:astro-ph/0501171}.

\bibitem[{\citenamefont{Simon et~al.}(2005)\citenamefont{Simon, Verde, and
  Jimenez}}]{Simon:2004tf}
\bibinfo{author}{\bibfnamefont{J.}~\bibnamefont{Simon}},
  \bibinfo{author}{\bibfnamefont{L.}~\bibnamefont{Verde}}, \bibnamefont{and}
  \bibinfo{author}{\bibfnamefont{R.}~\bibnamefont{Jimenez}},
  \bibinfo{journal}{Phys. Rev.} \textbf{\bibinfo{volume}{D71}},
  \bibinfo{pages}{123001} (\bibinfo{year}{2005}),
  \eprint{arXiv:astro-ph/0412269}.

\bibitem[{\citenamefont{Samushia and Ratra}(2006)}]{Samushia:2006fx}
\bibinfo{author}{\bibfnamefont{L.}~\bibnamefont{Samushia}} \bibnamefont{and}
  \bibinfo{author}{\bibfnamefont{B.}~\bibnamefont{Ratra}},
  \bibinfo{journal}{Astrophys. J.} \textbf{\bibinfo{volume}{650}},
  \bibinfo{pages}{L5} (\bibinfo{year}{2006}), \eprint{arXiv:astro-ph/0607301}.

\bibitem[{\citenamefont{Wei and Zhang}(2007)}]{Wei:2006ut}
\bibinfo{author}{\bibfnamefont{H.}~\bibnamefont{Wei}} \bibnamefont{and}
  \bibinfo{author}{\bibfnamefont{S.~N.} \bibnamefont{Zhang}},
  \bibinfo{journal}{Phys. Lett.} \textbf{\bibinfo{volume}{B644}},
  \bibinfo{pages}{7} (\bibinfo{year}{2007}), \eprint{arXiv:astro-ph/0609597}.

\bibitem[{\citenamefont{Steinhardt et~al.}(1999)\citenamefont{Steinhardt, Wang,
  and Zlatev}}]{Steinhardt:1999nw}
\bibinfo{author}{\bibfnamefont{P.~J.} \bibnamefont{Steinhardt}},
  \bibinfo{author}{\bibfnamefont{L.-M.} \bibnamefont{Wang}}, \bibnamefont{and}
  \bibinfo{author}{\bibfnamefont{I.}~\bibnamefont{Zlatev}},
  \bibinfo{journal}{Phys. Rev.} \textbf{\bibinfo{volume}{D59}},
  \bibinfo{pages}{123504} (\bibinfo{year}{1999}),
  \eprint{arXiv:astro-ph/9812313}.

\bibitem[{\citenamefont{Damour}(2001)}]{Damour:2001fn}
\bibinfo{author}{\bibfnamefont{T.}~\bibnamefont{Damour}},
  \emph{\bibinfo{title}{Questioning the equivalence principle}}
  (\bibinfo{year}{2001}), \eprint{arXiv:gr-qc/0109063}.

\bibitem[{\citenamefont{Su et~al.}(1994)\citenamefont{Su, Heckel, Adelberger,
  Gundlach, Harris, Smith, and Swanson}}]{Su:1994gu}
\bibinfo{author}{\bibfnamefont{Y.}~\bibnamefont{Su}},
  \bibinfo{author}{\bibfnamefont{B.}~\bibnamefont{Heckel}},
  \bibinfo{author}{\bibfnamefont{E.}~\bibnamefont{Adelberger}},
  \bibinfo{author}{\bibfnamefont{J.}~\bibnamefont{Gundlach}},
  \bibinfo{author}{\bibfnamefont{M.}~\bibnamefont{Harris}},
  \bibinfo{author}{\bibfnamefont{G.}~\bibnamefont{Smith}}, \bibnamefont{and}
  \bibinfo{author}{\bibfnamefont{H.}~\bibnamefont{Swanson}},
  \bibinfo{journal}{Phys. Rev.} \textbf{\bibinfo{volume}{D50}},
  \bibinfo{pages}{3614} (\bibinfo{year}{1994}).

\bibitem[{\citenamefont{Damour}(1996)}]{Damour:1996xt}
\bibinfo{author}{\bibfnamefont{T.}~\bibnamefont{Damour}},
  \bibinfo{journal}{Class. Quant. Grav.} \textbf{\bibinfo{volume}{13}},
  \bibinfo{pages}{A33} (\bibinfo{year}{1996}), \eprint{arXiv:gr-qc/9606080}.

\bibitem[{\citenamefont{Fuzfa and Alimi}(2006)}]{Fuzfa:2006pn}
\bibinfo{author}{\bibfnamefont{A.}~\bibnamefont{Fuzfa}} \bibnamefont{and}
  \bibinfo{author}{\bibfnamefont{J.~M.} \bibnamefont{Alimi}},
  \bibinfo{journal}{Phys. Rev. Lett.} \textbf{\bibinfo{volume}{97}},
  \bibinfo{pages}{061301} (\bibinfo{year}{2006}),
  \eprint{arXiv:astro-ph/0604517}.

\bibitem[{\citenamefont{Dickey et~al.}(1994)\citenamefont{Dickey, Bender,
  Faller, Newhall, Ricklefs, Ries, Shelus, Veillet, Whipple, Wiant
  et~al.}}]{Dickey:1994}
\bibinfo{author}{\bibfnamefont{J.~O.} \bibnamefont{Dickey}},
  \bibinfo{author}{\bibfnamefont{P.~L.} \bibnamefont{Bender}},
  \bibinfo{author}{\bibfnamefont{J.~E.} \bibnamefont{Faller}},
  \bibnamefont{et~al.}, \bibinfo{journal}{Science}
  \textbf{\bibinfo{volume}{265}}, \bibinfo{pages}{482} (\bibinfo{year}{1994}).

\bibitem[{\citenamefont{Arzoumanian}(2002)}]{Arzoumanian:2002xu}
\bibinfo{author}{\bibfnamefont{Z.}~\bibnamefont{Arzoumanian}},
  \emph{\bibinfo{title}{Improved bounds on violation of the strong equivalence
  principle}} (\bibinfo{year}{2002}), \eprint{arXiv:astro-ph/0212180}.

\bibitem[{\citenamefont{Lammerzahl}(2004)}]{Lammerzahl:2004tc}
\bibinfo{author}{\bibfnamefont{C.}~\bibnamefont{Lammerzahl}},
  \emph{\bibinfo{title}{General relativity in space and sensitive tests of the
  equivalence principle}} (\bibinfo{year}{2004}), \eprint{arXiv:gr-qc/0402122}.

\end{thebibliography}
\end{document}